\documentclass{article}

\usepackage{arxiv}

\usepackage[utf8]{inputenc} 
\usepackage[T1]{fontenc}    
\usepackage{hyperref}       
\usepackage{url}            
\usepackage{booktabs}       
\usepackage{amsfonts}       
\usepackage{nicefrac}       
\usepackage{microtype}      
\usepackage{lipsum}
\usepackage{graphicx}
\graphicspath{ {./images/} }

\usepackage{subcaption}

\title{Simulation, Model Checking, and Execution of Activity Models}

\author{
Abdurrahman Alshareef \\
Information Systems Department\\
College of Computer and Information Sciences\\
King Saud University\\
Riyadh 11451, Saudi Arabia\\
\texttt{ashareef@ksu.edu.sa}\\
\And
Hessam S. Sarjoughian \\
Arizona Center for Integrative M\&S\\
Schl. of Comp., Info., \& Deci. Syst. Engr.\\
Arizona State University\\
699 South Mill Avenue\\
Tempe, AZ, 85281, USA\\
\texttt{sarjoughian@asu.edu}\\

}

\begin{document}
\maketitle
\begin{abstract}
This paper presents our findings for using activity modeling for simulation (validation), model checking (verification), and execution purposes. Each is needed to tackle system complexity and further research into behavioral modeling. We argue different models implicate different understandings and expectations. We emphasize some distinctions with demonstrations using the Discrete Event System Specification with an exemplary model. In particular, the continuous-time base in models helps observe some inherent limitations and strengths in acquiring each capability. The temporal characterization of input, output, and state, or the lack thereof, is at the core of developing behavioral specifications. We use DEVS to arrive at the capability of validating simulations for activity models. We use Constrained-DEVS for the verification of activity models. We show how some executions can be derived, whether from the specification itself or with considerations for simulation and model checking. 
\end{abstract}


\section{Introduction}
\label{sec:intro}

We begin by describing specific ways by which activity models and diagrams have been treated to endow them with some features that go beyond their common use as static specification artifacts to aid in implementation and testing. They tend to introduce capabilities such as execution,  model checking, and simulation to the created artifacts. Different issues may arise along the path of arriving at each capability. Therefore, clarifications with further research and demonstrative efforts offer insights to gain a better and new understanding of behaviors of dynamical systems.

In previous works (e.g., \cite{alshareef2018activity}), we have shown the importance of having a DEVS-based simulation of activities from the standpoint of execution and verification. We argued that the semantics of some activity nodes, control, in particular, encounter non-trivial temporal ordering. It is helpful to examine such structures in simulation with a more rigorous account of time for the execution and model checking. For example, more than one activation path of nodes within a single activity can pose fundamental difficulties due to parallelism across action and control nodes. Thus, it is helpful to employ simulation to identify and handle such limitations in complementary and distinctive manners.

In the following section, we will discuss each capability with backgrounds and related works in detail. Then, we will present the tool that we have developed to endow activity with more sophisticated simulation capability. We then exemplify the use of each capability for activity modeling with a simple but non-trivial multi-server system.

\section{Verification vs. Simulation vs. Execution of Activities}

Modeling formalisms have been useful for a variety of needs, among them validation and verification. Common examples of these formalisms are Petri nets \cite{murata1989petri}, Timed Automata (TA) \cite{alur2015principles}, answer set programming (ASP) \cite{lifschitz1999answer}, and DEVS \cite{zeigler2000theory}. Each can be used to dissect certain aspects of a system's states, inputs/outputs, and state/output transitions. However, there has not been a formal way that a complete representation of a system can take place except with a large degree of constriction or abstraction. Thus, it becomes necessary to use a variety of formalisms for various needs in a complementary manner. Fundamental difficulties may arise, especially regarding heterogeneity; however, certain guarantees can be afforded if some analysis effort takes place. This fact is behind the high cost of developing models that conform to formal methods. Less formal approaches such as Unified Modeling Language (UML) and System Modeling Language (SysML) with the benefit afforded in Model Driven Architecture (MDA) is appealing in-practice and the use of accompanying frameworks and tools. Environments such as Eclipse Modeling Framework \cite{steinberg2008emf} although key, they may not lend themselves to a desired degree of rigor \cite{fondement2013big}.

Despite advances in the degree of coverage for specifying structure and behavior, restricted formalisms such as Petri nets and DEVS continue to require extensions like time Petri nets \cite{berthomieu1991modeling}, high-level Petri nets \cite{jensen2012high}, Finite-Deterministic DEVS \cite{hwang2009reachability} and Constrained-DEVS \cite{gholami2017modeling}. Introducing a continuous-time base in the specification leads to an inherent difficulty in interpreting any classical computational model from a system-theoretic standpoint. This difficulty is evident due to the necessary treatment in such a setting for challenging aspects of heterogeneity, composability, discretization, multiple resolutions, and the likes.

\begin{figure}[htb]
\centering
\includegraphics[width=0.95\linewidth]{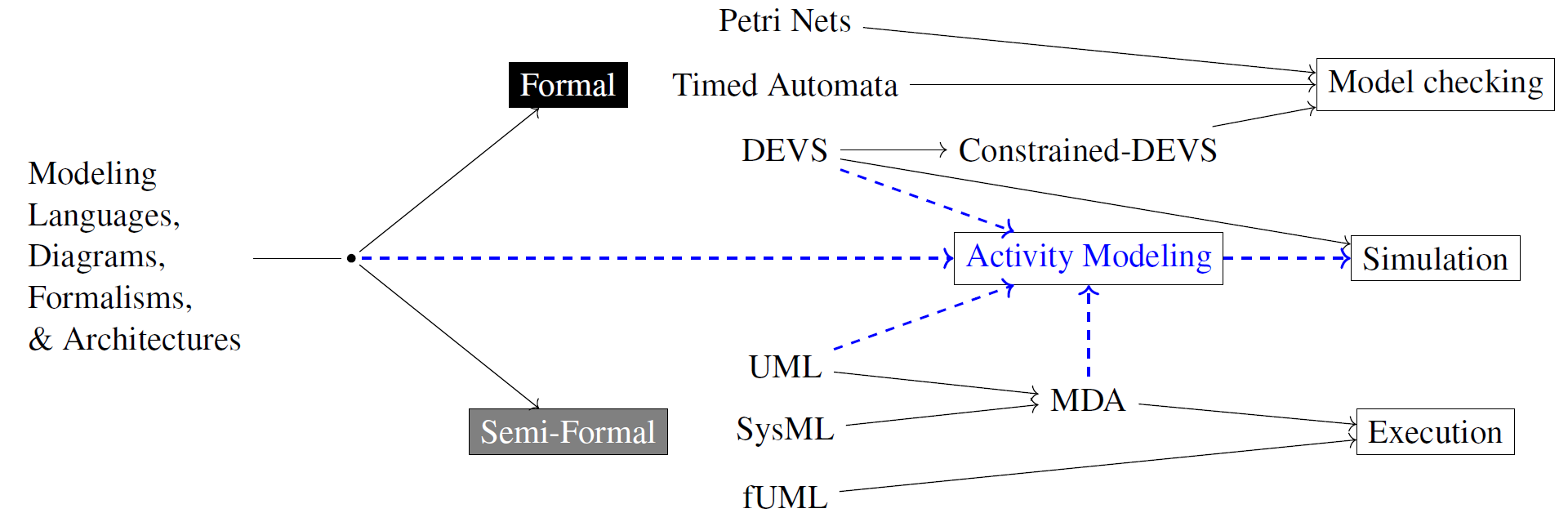}
\caption{Placing our contribution to activity modeling w.r.t select related works.}
\label{fig:related_works_map}
\end{figure}

Figure \ref{fig:related_works_map} highlights literature in this work, along with possible paths and research areas. Within this context, various efforts along the paths have been taken; however, others remain unexplored. Arriving at a particular execution is a capability of interest amongst researchers; others are verification and validation. The means of how such capabilities are delivered vary. Transformation, extension, and formalization capabilities are important to account for when model syntactic and execution semantic are inadequate. We will discuss in more detail some of the research on these three capabilities.
Nevertheless, it is a voyage in spaces with multiple paths to pursue and ideally arrive at useful methods with frameworks and tools for developing useful behavioral models.

Some formalisms account for specifications and are therefore used for modeling certain aspects of systems. The modeling efforts then translate to means in which verification of specific properties can take place under certain conditions. Some properties are reachability \cite{hwang2009reachability}, progress \cite{misra2001discipline} or liveness \cite{lamport1989simple}, maximality \cite{misra2001discipline}, and safety \cite{lamport1989simple,misra2001discipline,alur2015principles}. The path toward achieving verification of such properties involves creating counterpart representatives in formalisms like timed automata, Petri nets, or some extensions thereof, such as hybrid I/O automata \cite{lynch2003hybrid} or high-level Petri nets. Researchers have proposed different mappings to various formalisms or extensions thereof. The degree of coverage in these efforts also varies. Some of them account for basic elements \cite{rafe2008formal}, while others account for a wider set of constructs and definitions \cite{storrle2004semantics}, such as the ones defined in the activity metamodel \cite{uml2005unified}. For example, the latter includes detailed treatment for the semantics of various activity constructs such as executable node, control nodes, and various patterns of activity edges. The work concludes, however, with some critical remarks about the feasibility of aligning activities to Petri nets and vice versa. 

Considering the top part of the Figure~\ref{fig:related_works_map}, We account for the remarks on the semantics of the activity modeling by establishing the distinction between model checking and simulation as two different sought-after capabilities. The former can be supported by providing a counterpart to the path of Petri nets in the DEVS arena. Extension such as Finite-Deterministic DEVS (FD-DEVS) \cite{hwang2009reachability} and Constrained-DEVS \cite{gholami2017modeling} are developed for model checking. While the former depends on translating DEVS to TA, the latter is grounded on extending the DEVS formalism for model checking (verification) without DEVS to TA transformation. The DEVS-Suite modeling and simulation environments \cite{acims2020devs-suite} supports in a straightforward fashion both validation and verification with support for time series and behavior monitoring. This simulator also supports black-box unit testing that is useful for debugging and for validation \cite{DEVS-TestFrame-2020}.

On the other side (see the lower part of Figure \ref{fig:related_works_map}, some existing transformations and extensions suggest the notion of model execution and define semantics along the way during the model development and through proposed execution engines. Some of these approaches employ the MDA \cite{miller2003mda} approach and use methods such as model to text (M2T) and Query/View/Transformation (QVT). Tools developed for these methods enable code generation for producing code snippets and programs in target programming languages. An earlier work is executable UML \cite{mellor2002executable}. More recently, the foundational subset of UML (fUML) \cite{fuml2013} formalizes a subset of the UML using a set of actions with more elaborate semantic definitions and an execution model. It also proposes some mappings to the programming language Java. Such standardization efforts led to the development of execution engines for the UML activity diagram such as Moka \cite{Foundation2016}. The mapping is then drawn from the introduced specialization of activity elements (e.g., \textit{Read Self Action}) to their counterpart in the target programming language (e.g., \textit{this} in Java ). From a high-level point of view, the relationship between modeling constructs and their code counterpart in these frameworks is apparently one-to-one.

Some others works are aimed at overcoming challenges such as using a large graphical notation to represent a relatively simple procedure. Among them is the action language for foundational UML (Alf), a textual language to represent fUML \cite{francesco2017fUML}, which underscore the difficulty of automatically generating concrete models from fUML artifacts. Other approaches employ the MDA metamodeling, profiling, and other extension mechanisms to compensate for the UML and SysML with more concrete details of implementations. One goal is to equip them with better inclination to simulation \cite{Foures2012} or execution \cite{mayerhofer2013xmof}. We thoroughly examined these studies on various occasions, especially the ones that used DEVS, such as the efforts by \cite{Nikolaidou2008}, \cite{risco2009eudevs}, \cite{cetinkaya2011mdd4ms}, and \cite{kapos2019declarative}. In recent work \cite{alshareef2020Activity_DEVS_vs_fUML}, we highlight differences between treating activities from a DEVS standpoint as opposed to fUML.

Transformation is used for DEVS, on the one hand, and the UML and SysML using MDA, on the other (e.g., \cite{Yonglin2009}, \cite{sarjoughian2012emf}, \cite{mittal2013netcentric}). These studies attempted to holistically look into the problem of a potential mismatch between the formal and semi-formal specifications. The frameworks developed for MDA deliver some benefits for transformation from a primarily structural vantage point, and certain mappings are selective, leaving many details to be abstract given that every metamodel is inherently incomplete. Various adaptations and implementations take place at concrete layers to compensate and complement as much as possible for their counterpart representations at multiple higher layers. Lower level implementations are necessary to arrive at some artifacts that facilitate simulation or execution. However, the correspondence between formal and semi-formal specification remains mostly unfulfilled, especially regarding semantic definitions.

In this work, we attempt to use simulations for realizing, to some degree, the behavioral specification of certain types of models. We examine an inclusive subset of activity constructs while considering their syntax and semantics. From a modeling perspective, we employ the notion of the model as defined in system-theoretic formalisms for discrete-event or discrete-time model specifications. Models of such a nature encounter a more rigorous and explicit specification of their time base, I/O sets, state sets, and I/O and state segments. The result is models that benefit from basic definitions in general modeling languages and semi-formal methods. Yet, such models lend themselves to the mathematical discrete event system specification (i.e., the DEVS) and its abstract execution protocol.

\begin{figure}[htb]
\centering
\includegraphics[width=0.95\linewidth]{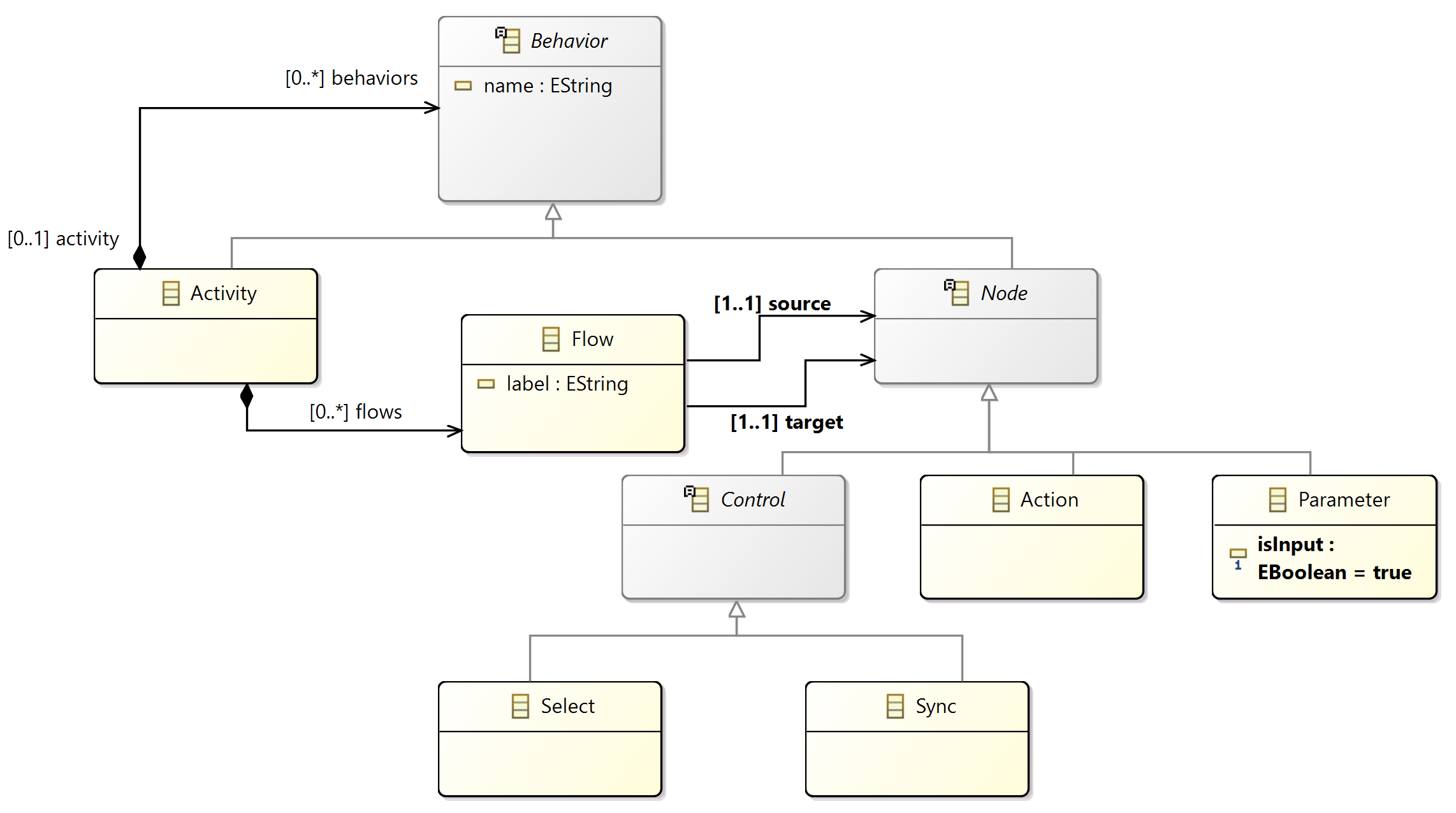}
\caption{A metamodel for hierarchical activities developed using Ecore.}
\label{fig:activity_ecore}
\end{figure}

\section{Activity Modeling Tool Demonstration}
\label{sec:demo_activity_tool}

The development of the tool starts by creating an Ecore model to account for the activity metamodel to support the creation of activity models in a hierarchical fashion. The hierarchy is accounted for in the composition relationship (see Figure \ref{fig:activity_ecore}) between \textit{Behavior} and \textit{Activity} elements. This model also shows the other \textit{EClass}, \textit{EReference}, and \textit{EAttribute} elements that are used to facilitate the model creation process.

\begin{figure}[!tbp]
  \centering
  \begin{subfigure}[b]{0.8\textwidth}
  \centering
    \includegraphics[width=0.7\textwidth]{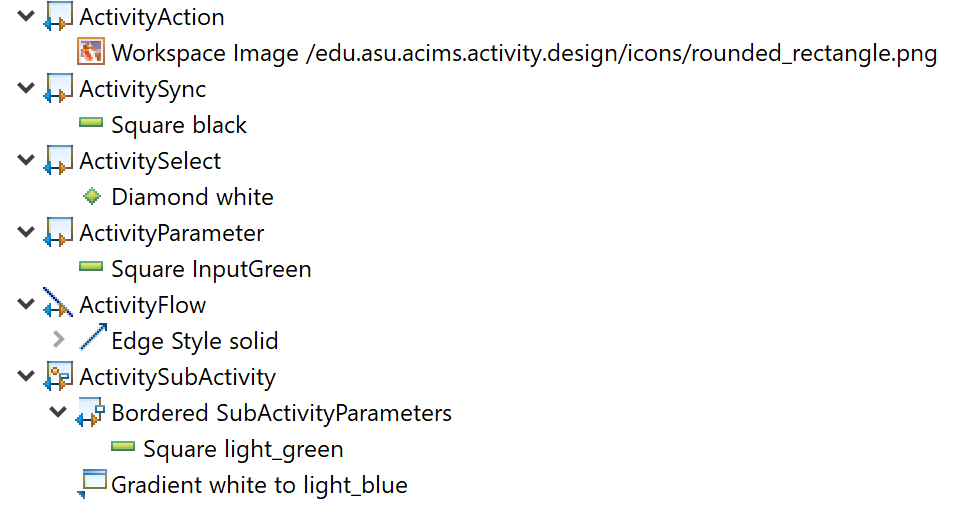}
    \caption{Viewpoint for different nodes creation with their designated geometrical shapes.}
    \label{fig:sirius_node_creation}
  \end{subfigure}
  \hfill
  \begin{subfigure}[b]{0.8\textwidth}
    \includegraphics[width=1\textwidth]{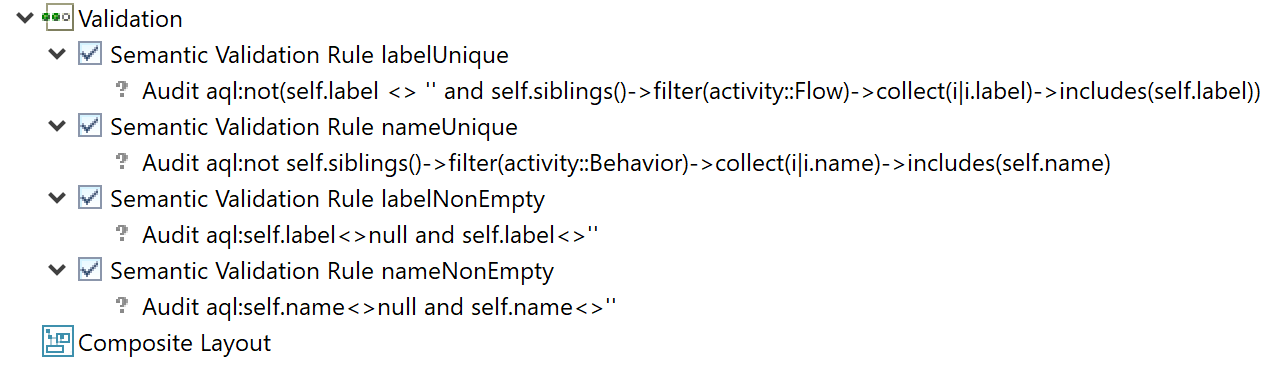}
    \caption{Viewpoint for validation such as ensuring label name uniqueness and setting.}
    \label{fig:sirius_validation}
  \end{subfigure}
  
  \caption{Viewpoint specifications in Sirius.}
  \label{fig:sirius_viewpoint}
\end{figure}

The graphical properties are defined using Viewpoint Specification in Sirius \cite{Sirius2018}. Figure \ref{fig:sirius_node_creation} shows the part for defining the geometric shapes by which the model artifacts are to visualize. For example, a rectangle with rounded corners refers to the action node. Similarly, other shapes are defined for other nodes in addition to the edge to represent activity flow. Figure \ref{fig:sirius_validation} defines some rules to issue a warning or error message to notify the modelers. Examples of rules are shown in the screen-shot, such as the label name uniqueness and setting. The geometric shapes are associated with definitions from the domain model (Ecore). The section offers further capabilities to manipulate other properties, such as the context of the created element. It also offers to modify the instance model upon changes. The rules need to be defined within the viewpoint specification file using languages such as Object Constraint Language (OCL) or Acceleo Query Language (AQL). An example is the deletion of flows upon the deletion of their corresponding nodes, implemented in the tool.

\begin{figure}[!tbp]
  \centering
  \begin{subfigure}[b]{0.7\textwidth}
  \centering
    \includegraphics[width=1\textwidth]{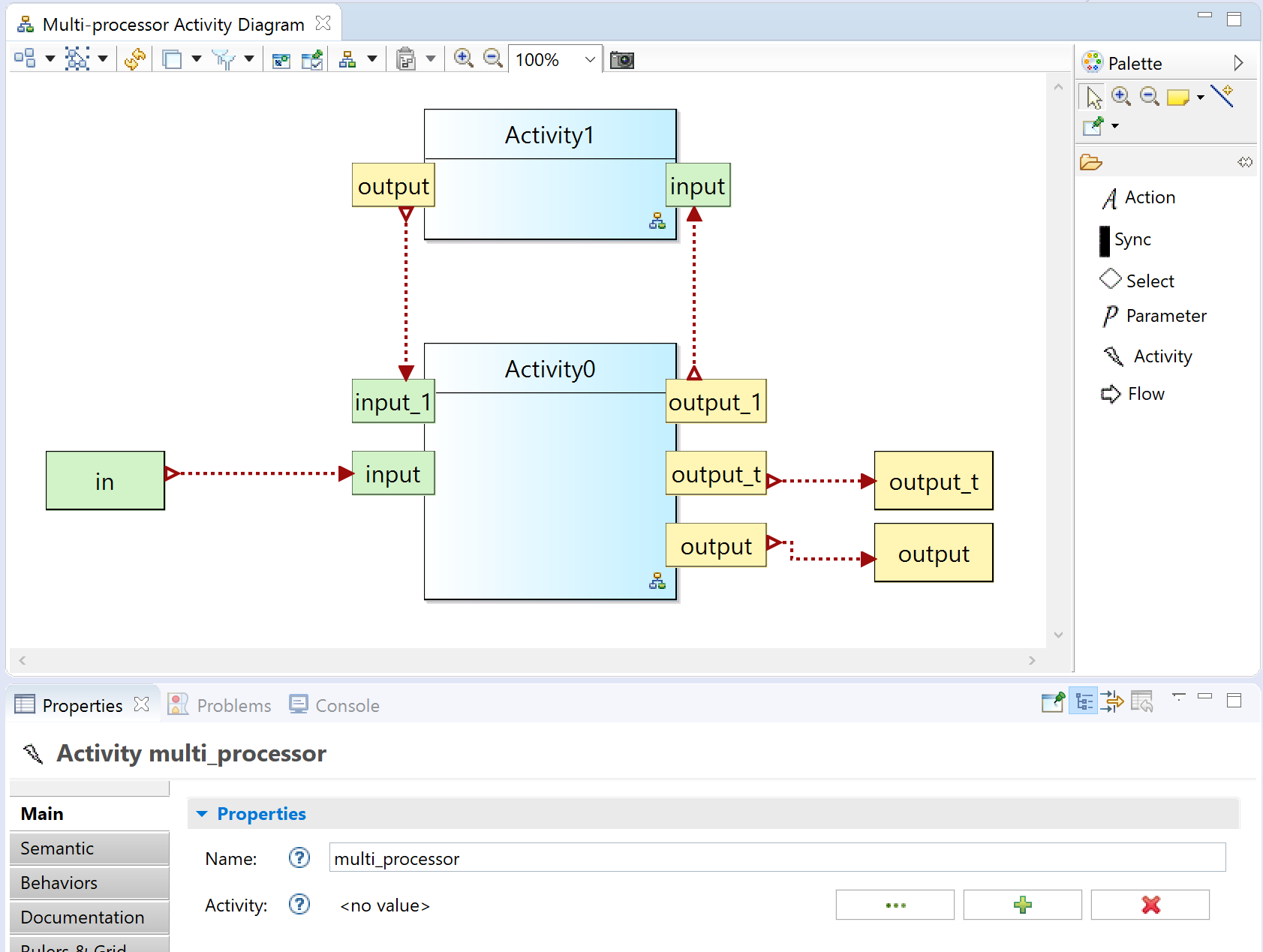}
    \caption{The screen-shot shows developing the main activity diagram for the multi-server with hierarchical construction as described in \cite{alshareef2018metamodeling}. In addition to the canvas, the palette and the properties view for the activity are shown.}
    \label{fig:multi-proc-activity}
  \end{subfigure}
  \begin{subfigure}[b]{0.7\textwidth}
    \includegraphics[width=1\textwidth]{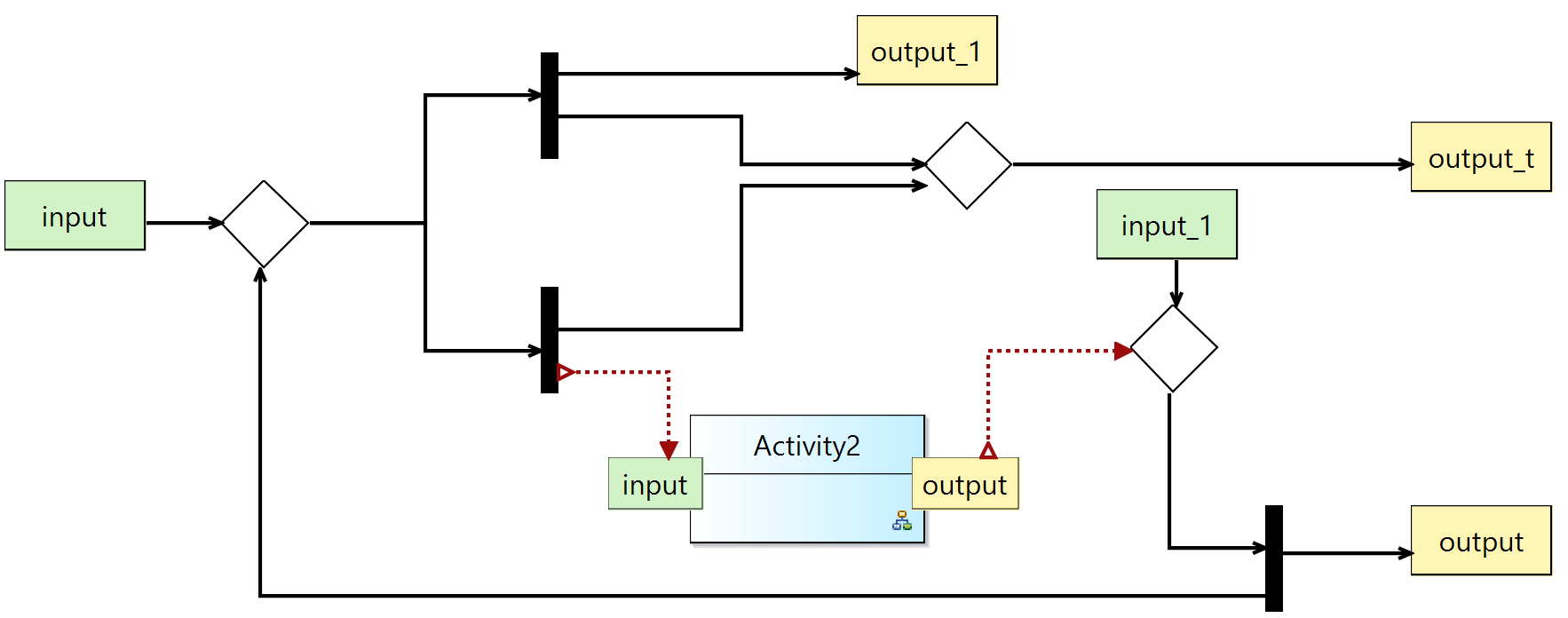}
    \caption{The corresponding diagram for activity 0.}
    \label{fig:activity0_diagram}
  \end{subfigure}
  \begin{subfigure}[b]{0.7\textwidth}
    \includegraphics[width=1\textwidth]{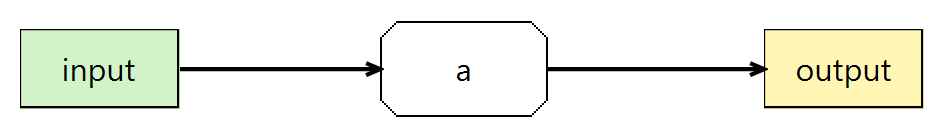}
    \caption{The corresponding diagram for activity 1 and 2.}
    \label{fig:activity1_diagram}
  \end{subfigure}
  \hfill
  
  \caption{Modeling multi-server activity in the developed activity modeling tool.}
  \label{fig:multi-server-activity}
\end{figure}

\begin{figure}[htb]
\centering
\includegraphics[width=0.95\linewidth]{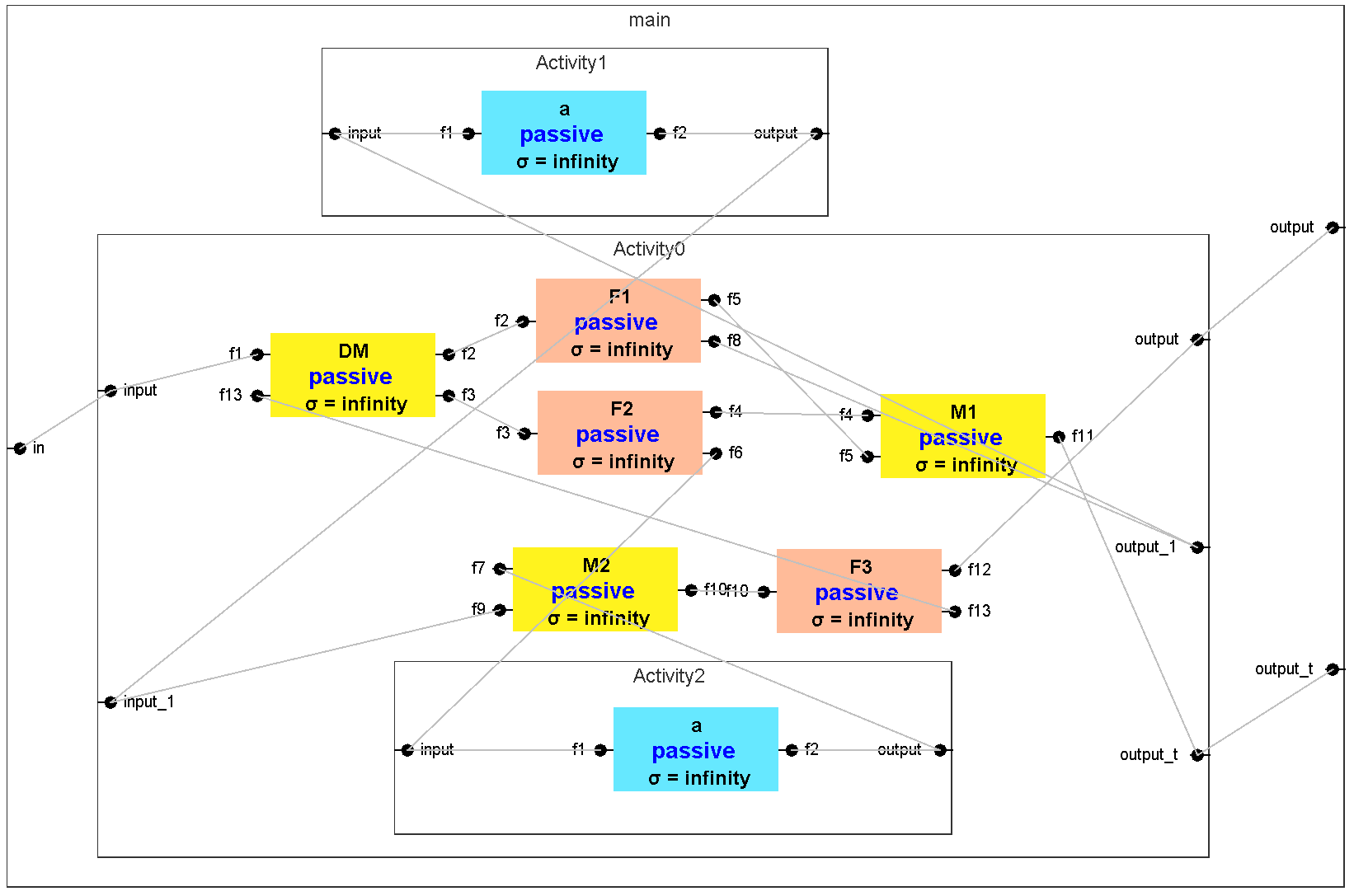}
\caption{Simulation view of a multi-server system after the code generation for DEVS-Suite simulator.}
\label{fig:multi-server-activity_sim_view}
\end{figure}

The tool then can be used to create diagrams such as the one for the multi-server activity (Figure \ref{fig:multi-server-activity}). Figure \ref{fig:multi-proc-activity} shows the creation of the main activity along with the tool palette. In the subsequent Figure \ref{fig:activity0_diagram} and \ref{fig:activity1_diagram}, we only show a screen-shot of the canvas. After developing these activities, the code generation process can take place. The code generators are implemented using Acceleo and the result in the set of models as Java files that are necessary for the simulation to take place in the DEVS-Suite simulator, as shown in Figure \ref{fig:multi-server-activity_sim_view}. The tool can generate code for Markov-DEVS models and simulated in MS4 Me \cite{ms4me}.

\section{DEVS-based Simulation of Activities}

The atomic model in DEVS \cite{zeigler2000theory} is a structure 
$ <X,S,Y,\delta_{int},\delta_{ext},\lambda,ta> $ where $ X $ is the set of input values, $ S $ is the set of states, $ Y $ is the set of output values, $ \delta_{int}:S\rightarrow S $ is the internal transition function, $ \delta_{ext}:S \times X\rightarrow S $ is the external transition function. $ Q $ is the total state set defined with the time elapsed since last transition. 
$ \lambda:S\rightarrow Y $ is the output function and $ ta:S\rightarrow \mathbb{R}_{0,\infty}^+ $ is the set of real positives numbers with 0 and $\infty$.

The role of the activity model is primarily to serve as an intermediary stage between the behavioral specification as defined in the DEVS formalism and their implementation in some concrete simulator such as DEVS-Suite and MS4 Me. The EMF-DEVS \cite{sarjoughian2012emf} is proposed to provide a modeling means for the Parallel DEVS based on metamodeling principles and the MDA. It uses the Eclipse Modeling Framework (EMF) to provide a suitable means for creating DEVS metamodels and exploit the validation infrastructure to define a set of constraints and apply them while accounting for both domain-neutral and domain-specific needs. Activity-based DEVS is built upon the DEVS formalism along with MDA framework and metamodeling principles in general. The goal is to examine arriving at richer behavioral specifications for discrete event models through intermediary stages. 

In the beginning, the meta-layers hierarchy is defined for DEVS behavioral specifications. Details of the M2 as a meta-layer, M1 as a modeling layer, and M0 for modeling instantiation are crafted to account for as many elements as possible without sacrificing rigor nor expressiveness. The issue resides in having models developed across different stages concerning their abstraction and hierarchy. The transition from mathematical to UML and from UML to some realizations such as EMF can be challenging. Multiple representations of a model at higher layers pose challenges at the implementation level, whether for verification or simulation purposes. Complexity and scale traits \cite{sarjoughian2017restraining} poses challenges with models of non-trivial dynamics. The depicted metamodel in \cite{sarjoughian2015behavioral} shows an EMF realization of meta-behavior concepts of the atomic DEVS model.

Due to the fact that MDA and UML are both semi-formal approaches, it remains difficult to obtain complete implementations of system elements despite their role in the taming of model structurally and behaviorally. Some behavioral constructs in the UML provide a suitable abstraction for elements at the low-level implementations. They may potentially provide some basis for richer specifications if they are used in a complementary manner and equipped with rigorous state and time definitions such as in DEVS.

Thus, we treat activities and actions from a DEVS standpoint, including their structural and behavioral properties, with a mapping proposed to facilitate the process of understanding the bridging points between activities, including their nodes and edges, and the DEVS formalism. The mapping includes the general constructs, such as actions and and control nodes, and more concrete constructs thereof, such as reading actions and specific fork nodes.

The activity node, which is an abstract node element in activities, is generally specified by an atomic model while remaining extensible for further specializations. The resemblance of the action and the atomic model is captured via the mapping. The notion of action provides the basis for arriving at more specific behavioral specifications while the DEVS formalism provides the basis for arriving at precise semantics \cite{alshareef2017toward}. The control nodes in activities offer different mechanisms to handle the flow in the model. Therefore, they can be closely examined through the lens of the DEVS formalism and Parallel DEVS in particular. The \textit{fork} and \textit{join} nodes are examined given their synchronizing properties. The other set of control nodes, which are \textit{decision} and \textit{merge} nodes, are also examined given their flow selection mechanisms. For both sets, a wide range of semantics are examined, especially the one with parallelism traits (further discussion is given in \cite{alshareef2018parallelism}).

\section{Verification of Activities}

\begin{figure}[ht]
\centering
\begin{subfigure}[b]{0.48\textwidth}
    \centering
    \includegraphics[width=0.9\linewidth]{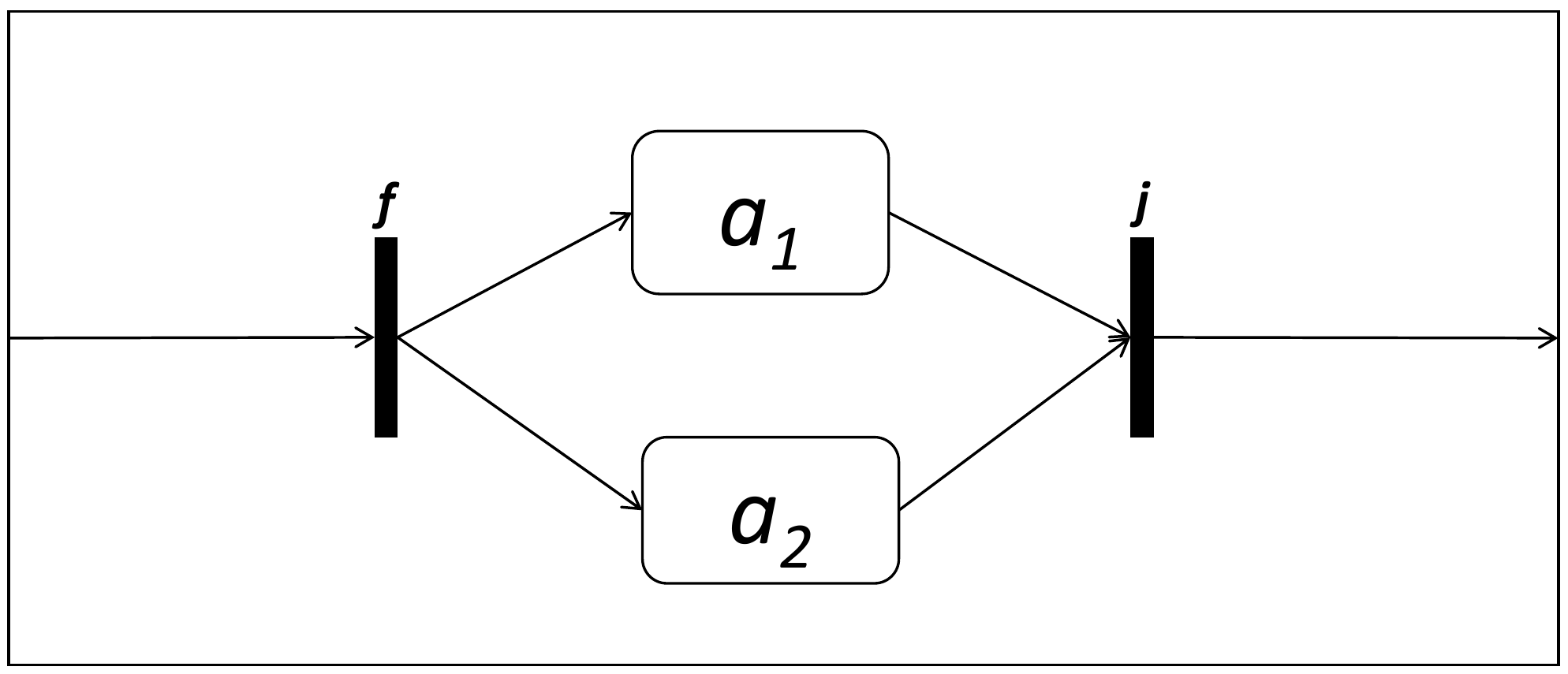}
   \caption{An activity abstraction for the divide and conquer.}
   \label{fig:DandC_activity}
\end{subfigure}
\hfill
\begin{subfigure}[b]{0.48\textwidth}
    \centering
   \includegraphics[width=0.98\linewidth]{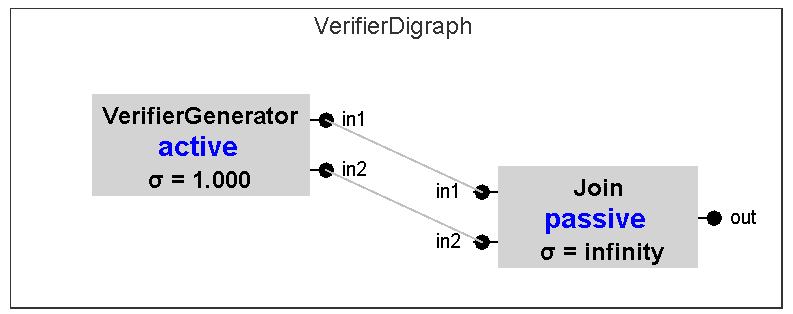}
   \caption{The verification digraph for the \textit{Join} element.}
   \label{fig:verifierDigraph}
\end{subfigure}

  \begin{subfigure}[b]{0.9\textwidth}
    \includegraphics[width=0.95\textwidth]{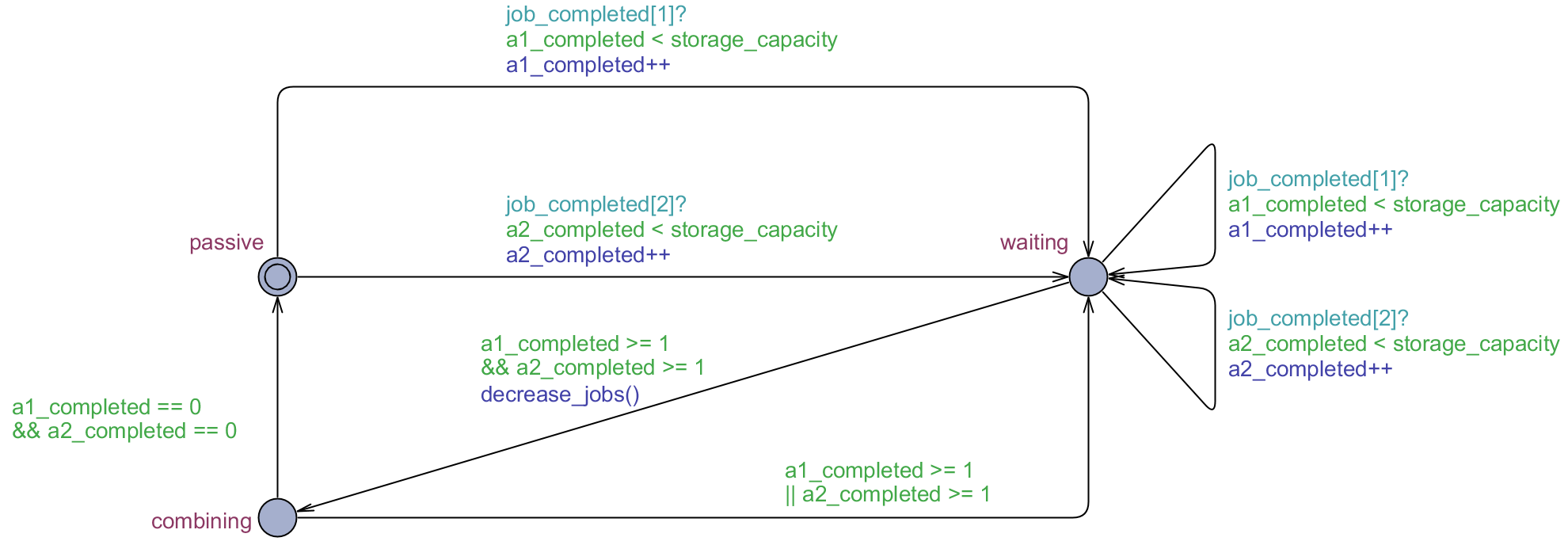}
    \caption{The TA representation of the \textit{join} element in UPPAAL.}
    \label{fig:uppaalJoin}
  \end{subfigure}
  
\caption{The modeling and verification of the exemplary processing architecture archetype.}
\label{fig:modelingVerificationOfDC}
\end{figure}

Previously, we have presented how to perform the verification using ASP \cite{alshareef2018model}. It is also possible to use TA for a similar purpose. Here, we demonstrate the verification of activities with Constrained-DEVS \cite{gholami2017modeling} and TA using \cite{uppaal4}. The action-level specification of activities is equipped with the notion of the state as defined in DEVS. Therefore, Constrained-DEVS as a candidate for performing verification is defined to represent the state-based behavior at the level of action or control node. It is then examined after being coupled with other models into a system under study. DEVS-Suite incorporates this capability and therefore is used to verify certain properties of models for the divide and conquer multi-processing regime. The activity in Figure \ref{fig:DandC_activity} is presented in previous work along with a demonstration of the use of DEVS-Suite simulator in observing its corresponding behavior . Here, we will use constrained DEVS to verify certain aspects of this activity.

Four Constrained-DEVS components are defined and created to develop the divide and conquer multi-processing architecture. A generator, which is defined by the verification engine, fork, action, and join models are created to mimic and restrict the behavior of their counterparts in the models defined for DEVS-Suite and for the divide architecture archetype. Figure \ref{fig:verifierDigraph} shows the verifier for the \textit{join} element. The system is then composed of five models, generator, fork, two instances of action, and join. The state set is restricted to be finite in space to enable the verification process. Each variable, such as the phase and the storage capacity, is restricted to have a finite range. Finally, a transducer is defined to check for the functional properties that are subject to the verification.

With respect to verifier, the deadlock property is defined $\phi$ to be checked after constraining the state space. It is checked for multiple instantiation scenarios of processes instances of the action template. One scenario is defined where $action_1$ is instantiated with processing time 1 and $action_2$ is instantiated with processing time 6. Other instantiation scenarios are defined and then checked for two formulae using three properties. The result of checking is shown in Table~\ref{tab:property_check} after composing all templates according to the configuration in Figure \ref{fig:DandC_activity}. $action_2$ is instantiated with different settings of the processing times. These settings are 4, 5, and 6. For each scenario, the defined formulae are checked. The deadlock property is defined $\phi$. The second property is defined $\psi$, which refers to the global clock being greater than 5. Finally, the property $\rho$ is defined for the $action_2$ to state that it is in phase busy. The second formula in the table $\forall \Box \ \psi \Rightarrow \rho$ means that the $action_2$ remains \textit{busy} when after 5 time units. The checking of this formula when the processing time for this action is 6 or 8 is consistent with the obtained state trajectory generated by the simulator.

\begin{table}
 \caption{Checking different path formulae in different processing time settings for the second action within divide and conquer architecture. The generation period is set to 5 and the processing time for the first action is set to 1. The deadlock happens due to infinite waiting by the \textit{join} model because of the $action_2$ being in phase \textit{busy} indefinitely, and an imposed limitation on the storage capacity.}
  \centering
  \begin{tabular}{llll}
    \toprule
    Processing time in $processor_2$   & 4     & 5 & 6 \\
    \midrule
    $\forall \Box \ \neg \phi$ & Satisfied & Not satisfied & Not satisfied
        \\
    $\forall \Box \ \psi \Rightarrow \rho$ & Not satisfied & Not satisfied & Satisfied      
    \\
    \bottomrule
  \end{tabular}
  \label{tab:property_check}
\end{table}

\section{Conclusion}

Researchers endeavor to enrich the model development process by incorporating fundamental capabilities (e.g., simulation, verification, and execution) in an integrative fashion. A variety of techniques and tools have been developed based on different understandings and expectations. A perspective on this line of research is proposed and detailed in this paper. A variety of frameworks and tools supports various modeling concepts and capabilities key to specifying behaviors of complex systems, particularly those that are time-driven. In some, one capability like execution can be the primary,  while others are serving specific needs. For example, the DEVS-Suite simulator is suited for simulating parallel DEVS models. It can also be used for verification of Constrained DEVS models. Similarly, UPPAAL is suited for verification and also lends itself for simulating finite-state automata that have clocks.

Our work focuses on developing a DEVS counterpart model to carry out the activity model semantics in a more precise manner for the major activity constructs, which are action and control nodes. We model control nodes into two major atomic models, \textit{Sync} and \textit{Select}. The \textit{Sync} atomic model is developed to represent the specification of the \textit{fork} and \textit{join} nodes. And the \textit{Select} atomic model is developed to represent the specification for the \textit{decision} and \textit{merge} nodes. We manage to obtain simulations that conform to the semantics of the activity diagram. Moreover, we account for some semantics that is absent in some specifications of the UML activities such as input ports for control nodes or the arbitrary processing upon the arrival of multiple inputs. We show that our approach enriches the activity modeling process by making them subject to nuanced questions and complex experiments as encountered in the theory of modeling and simulation with the use of experimental frames. More details of our work on activity modeling including  examples and experiments \cite{alshareef2019diss} are to be presented in future work.

\bibliographystyle{unsrt}  
\bibliography{references}  






\end{document}